\documentclass[preprint,showpacs,showkeys,preprintnumbers,amsmath,amssymb,prb]{revtex4}

\usepackage{graphicx}
\usepackage{float}

\begin{document}

\preprint{Winterlik et al, ZrNi$_2$Ga}

\title{A Ni-based Superconductor: the Heusler Compound ZrNi$_2$Ga.}  

\author{ J{\"u}rgen~Winterlik, Gerhard~H.~Fecher and Claudia~Felser}
\email{felser@uni-mainz.de}
\affiliation{Institut f\"ur Anorganische und Analytische Chemie,
             Johannes Gutenberg - Universit\"at, 55099 Mainz, Germany.}

\author{Martin Jourdan}
\affiliation{Institut f\"ur Physik, Johannes Gutenberg - Universit\"at, 55128 Mainz, Germany.}

\author{Kai Grube}
\affiliation{Forschungszentrum Karlsruhe, Institut f\"ur
Festk\"orperphysik,
P.O. Box 3640, 76021 Karlsruhe, Germany}

\author{Fr\'{e}d\'{e}ric Hardy and Hilbert von L\"ohneysen}
\affiliation{Forschungszentrum Karlsruhe, Institut f\"ur Festk\"orperphysik,
             P.O. Box 3640, 76021 Karlsruhe, Germany,
             and Physikalisches Institut, Universit\"at Karlsruhe, 76128 Karlsruhe, Germany}

\author{K. L. Holman and R. J. Cava}
\affiliation{Department of Chemistry, Princeton University, Princeton, New Jersey 08544, USA}

\date{\today}

\begin{abstract}

This work reports on the novel Heusler superconductor ZrNi$_2$Ga. Compared
to other nickel-based superconductors with Heusler structure, ZrNi$_2$Ga exhibits
a relatively high superconducting transition temperature of $T_c=2.9$~K and 
an upper critical field of $\mu_0H_{c2}=1.5$~T. Electronic structure calculations
show that this relatively high T$_c$ is caused by a van~Hove singularity,
which leads to an enhanced density of states at the Fermi energy $N(\epsilon_F)$.
The van~Hove singularity originates from a higher order valence instability at
the $L$-point in the electronic structure. The enhanced $N(\epsilon_F)$
was confirmed by specific heat and susceptibility measurements. Although
many Heusler compounds are ferromagnetic, our measurements of ZrNi$_2$Ga 
indicate a paramagnetic state above $T_c$ and could not reveal any traces 
of magnetic order down to temperatures of at least 0.35~K. We investigated 
in detail the superconducting state with specific heat, magnetization, and 
resistivity measurements. The resulting data show the typical behavior of a
conventional, weakly coupled BCS (s-wave) superconductor.

\end{abstract}



\maketitle

\section{Introduction}

In the research area of spintronics applications, Heusler compounds have become 
of interest as half-metals, where due to the exchange splitting of 
the $d$-electron states, only electrons of one spin direction have a finite 
density of states at the Fermi level $N(\epsilon_F)$~\cite{KFF07,FFB07}. 
Up to the present, very few Heusler superconductors with the ideal formula
of AB$_2$C have been found. In 1982, the first Heusler superconductors were 
reported, each with a rare-earth metal in the B position~\cite{IJJ82}. Among
the Heusler superconductors, Pd-based compounds have attracted attention because
YPd$_2$Sn exhibits the highest yet recorded T$_c$ of 4.9~K~\cite{WHB83}.
Moreover, coexistence of superconductivity and antiferromagnetic order was
found in YbPd$_2$Sn~\cite{KDM85} and ErPd$_2$Sn~\cite{SHJ86}. A systematic 
investigation of Ni-based Heusler compounds seems to be worthwhile as nickel 
has many properties in common with palladium but tends more towards magnetic 
order due to the smaller hybridization of the $3d$-states. In fact, elementary
nickel is a ferromagnet. Thus, nickel-containing Heusler compounds with a high 
proportion of Ni are naively expected to show magnetic order rather than 
superconductivity. However, superconductivity of Ni-rich alloys NbNi$_2$C 
(C = Al, Ga, Sn) has been reported some time ago, with transition temperatures $T_c$
ranging from 1.54~K to the highest recorded transition temperature of a Ni-based
Heusler compound of 3.4~K in NbNi$_2$Sn~\cite{WHB83,WYM85}. In contrast to the 
two aforementioned Pd-based compounds these superconductors do not show indications 
of magnetic order. Currently there is a lot of excitement about the new high 
temperature superconductors based on FeAs~\cite{KHH06}. The superconductivity of
these compounds is related to two-dimensional layers of edge shared FeAs 
tetrahedrons~\cite{HCZ08}. These structure types can be understood as 
two-dimensional variants of the Heusler structure.

A clear understanding of the origin of superconductivity, magnetism, and their 
possible coexistence in Heusler compounds is still missing. To shed light on the 
relation between the electronic structure and the resulting ground state of AB$_2$C
Heusler compounds we searched for new Ni-based Heusler compounds with a high density 
of states (DOS) at $\epsilon_F$ close to the Stoner criterion for ferromagnetism. A 
possible route for increasing $N(\epsilon_F)$ is the use of saddle points in the
energy dispersion curves of the electronic structure. They lead to maxima in the 
DOS, so-called van~Hove singularities~\cite{vHo53}. In order to identify such
compounds, we have performed electronic structure calculations using \textit{ab initio} 
methods. In a simple approach following the Bardeen-Cooper-Schrieffer theory (BCS) 
and neglecting any magnetic order, we would expect that the superconducting 
transition temperature of such compounds increases with $N(\epsilon_F)$ according 
to $T_c\approx\Theta_D\exp(-1/V_0 N(\epsilon_F))$ if the Debye temperature $\Theta_D$
and the Cooper-pairing interaction V$_0$ are independent of $N(\epsilon_F)$. In 
fact, this van~Hove scenario, where a maximum in the DOS is ideally located at $\epsilon_F$, 
was used to explain the unusually high transition temperatures of the intermetallic
A15 superconductors~\cite{LaF66}. The correspondence between $T_c$ and the valence 
electron count is known as Matthias rule~\cite{Mat53}. According to this rule, the
high $T_c$ of the A15 compounds was related to electron concentrations of about 4.6
and 6.4 electrons per atom, leading to a maximum of the DOS at $\epsilon_F$~\cite{VIK82}.

On the basis of the van~Hove scenario, we already found superconductivity in two 
Heusler compounds with 27 electrons: ZrPd$_2$Al and HfPd$_2$Al~\cite{Fel01,WFF08}. 
Here, we report on the theoretical and experimental characterization of the new,
Ni-containing, superconducting Heusler compound ZrNi$_2$Ga. Additionally, electron-doped 
alloys Zr$_{1-x}$Nb$_x$Ni$_2$Ga were prepared and investigated to obtain information 
about the dependence of $T_c$ on the location of the van~Hove singularity.

\section{Experimental details}
\label{sec_ed}

Polycrystalline ingots of ZrNi$_2$Ga and electron-doped alloys Zr$_{1-x}$Nb$_x$Ni$_2$Ga 
were prepared by repeated arc melting of stoichiometric mixtures of the corresponding 
elements in an argon atmosphere at a pressure of 10$^{-4}$~mbar. Care was taken to 
avoid oxygen contamination. The samples were annealed afterward for 2~weeks at 1073~K 
in an evacuated quartz tube. After the annealing process, the samples were quenched 
in a mixture of ice and water to retain the desired $L2_{1}$ structure. The crystal
structure of ZrNi$_2$Ga was investigated using powder X-ray diffraction (XRD). The 
measurements were carried out using a Siemens D5000 with monochromatized Cu K$_\alpha$ 
radiation. 

The electrical resistance of a bar shaped sample was measured using a four-point probe
technique. The magnetization measurements below a temperature of 4~K were performed in 
a superconducting quantum interference device (SQUID, Quantum Design MPMS-XL-5). For 
higher temperatures, the magnetization was measured using a vibrating sample magnetometer 
(VSM option of a Quantum Design PPMS). The measured samples had a spherical shape with 
a mass of approximately 20~mg to 120~mg. In order to study the diamagnetic shielding,
the sample was initially cooled down to $T=1.8$~K without applying any magnetic field, i.e.,
zero-field cooled (ZFC). Then a field of $\mu_0H=2.5$~mT was applied, and the sample 
magnetization was recorded with increasing temperature. To determine the Meissner effect 
(flux expulsion) the sample was subsequently cooled and its magnetization measured in
the identical field, i.e., field cooled (FC). The field dependent magnetization of 
ZrNi$_2$Ga was measured at a temperature of 2~K. Finally, the normal-state susceptibility 
was measured at $\mu_0H=2$~T in a temperature range from 1.8~K to 300~K. Specific-heat
measurements were carried out at $0.35$~K~$<T<4$~K in magnetic fields of up to 5~T in 
a Quantum Design PPMS with a $^3$He option.

\section{\textit{Ab initio} calculations of the electronic and vibrational properties.}
\label{sec_cd}

The electronic and vibrational properties were calculated through the use of 
{\scshape Wien}2k \cite{SBM02} in combination with {\scshape Phonon} \cite{Par06}.
The electronic structure of ZrNi$_2$Ga was calculated by means of the full 
potential linearized augmented plane wave (FLAPW) method as implemented in
{\scshape Wien}2k provided by Blaha, Schwartz, and coworkers \cite{BSS90,BSM01,SBM02}.
The exchange-correlation functional was taken within the generalized gradient
approximation (GGA) in the parameterization of Perdew, Burke and Enzerhof
\cite{PBE96}. A $25\times25\times25$ point mesh was used as base for the 
integration in the cubic systems resulting in 455 $k$-points in the irreducible 
wedge of the Brillouin zone. The energy convergence criterion was set to 
$10^{-5}$~Ry and simultaneously the criterion for charge convergence to 
$10^{-3} e^{-}$. The muffin tin radii were set to 2.5~$a_{0B}$ 
($a_{0B}$~:=~Bohr's radius) for the transition metals as well as the main
group element. A volume optimization resulted in $a_{opt}=6.14$~\AA and a bulk
modulus of $B=156~GPa$ for the relaxed structure. This value is slightly larger
than the experimentally observed lattice parameter $a_{exp}$ (see below). The
results presented in the following are for the relaxed lattice parameter, no
noticeable changes are observed in the calculations using $a_{exp}$.

\begin{figure}[h]
\centering
\includegraphics[width=\columnwidth]{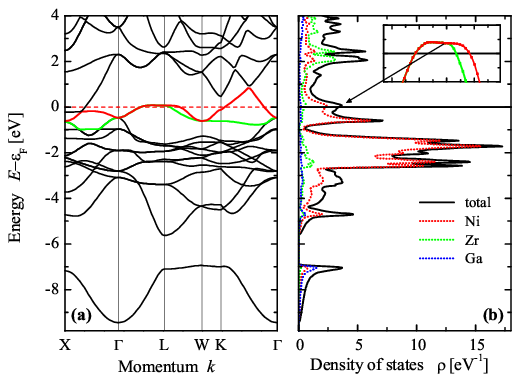}
\caption{Electronic structure of ZrNi$_2$Ga. 
         (a) displays the band structure and (b) the density of states.
         The inset in (b) shows the dispersion of the bands that cause
         the van Hove singularity at the $L$-point on an enlarged scale.}
\label{fig_spagdos}
\end{figure}

Figure~\ref{fig_spagdos} shows the results for the electronic structure from 
the \textit{ab initio} calculations. Typical for Heusler compounds is the low
lying hybridization gap at energies between 7~eV and 5.6~eV below the Fermi 
energy. This gap emerges from the strong interaction of the $s-p$ states at the
Ga atoms in O$_h$ symmetry with the eight surrounding Ni atoms. It explains the
structural stability of the compound.


More interesting are the bands close to the Fermi energy. In particular, 
the topmost valence band exhibits a van Hove singularity at the $L$-point 
only 70~meV above $\epsilon_F$. The result is a maximum of the density of 
states at the Fermi energy (see Figure~\ref{fig_spagdos}(b)). A closer 
inspection of those states reveals that the singularity at $L$ is a $S_2$-type
saddle point of the electronic structure with a twofold degeneracy. This 
degeneracy is removed along $LK$ or $LW$. For both bands, two of the second
derivatives $\left| \partial^2 E(k) / \partial k_i \partial k_j \right|_{k_e}$
of the dispersion $E(k)$ are $>0$ and one is $<0$ ($\Lambda$-direction) at
$k_e=(1/4,1/4,1/4)$.

\begin{figure}[h]
\centering
\includegraphics[width=\columnwidth]{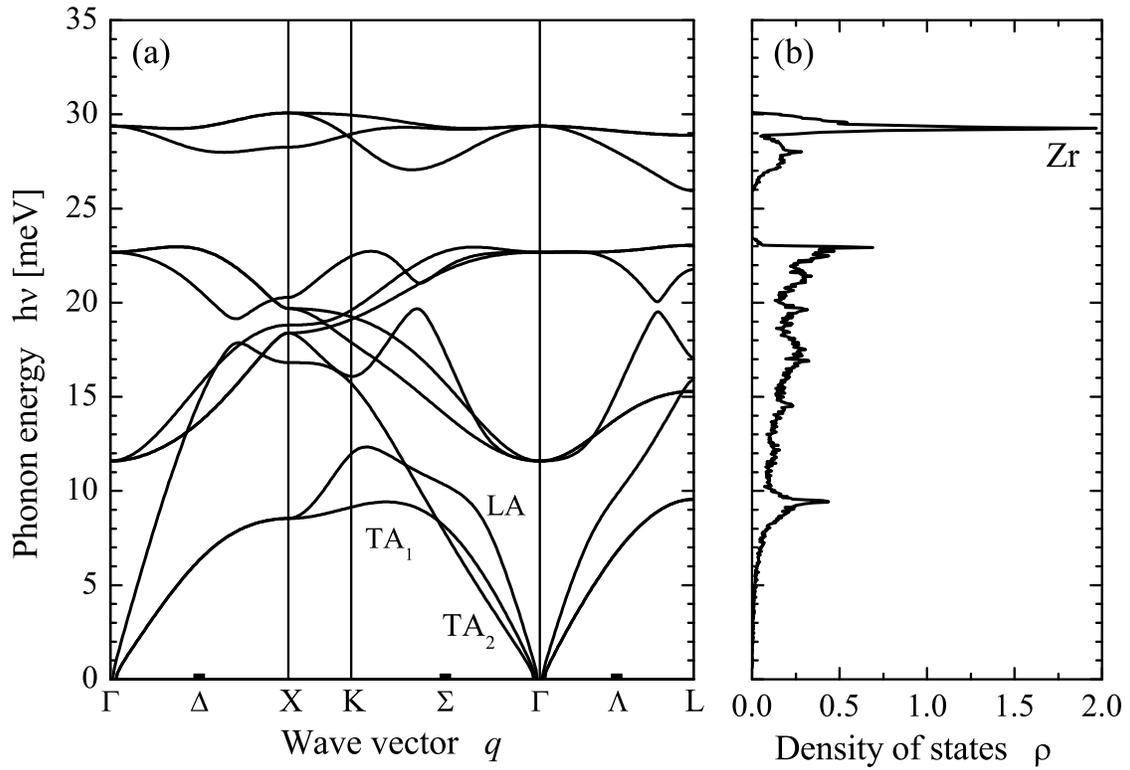}
\caption{The calculated vibrational spectrum of ZrNi$_2$Ga. 
         (a) displays the phonon dispersion and
         (b) the corresponding density of states.}
\label{fig:fig_phonon}
\end{figure}

Figure~\ref{fig:fig_phonon} shows the calculated phonon dispersion and phonon
density of states. The dispersion of the acoustic $LA$ and $TA_1$ modes is 
degenerate in the fourfold $\Delta$ direction as well as along $\Lambda$. 
This degeneracy is removed at the $K$-point and in the twofold $\Sigma$ 
direction. Instabilities in the form of soft-phonon modes, as are observed 
for several magnetic Ni-based Heusler compounds \cite{ZEn05,ZER05}, do not 
occur in the phonon dispersion relation of ZrNi$_2$Ga. This indicates the high
structural stability of the compound compared to the Ni-based Heusler shape memory 
alloys (for example Ni$_2$MnGa).


The high density of phonons at energies of about 30~meV is due to the
vibration of the rather heavy Zr atoms. These optical modes have no overlap
with the remainder of the phonon spectrum and appear as Einstein frequencies. 
In a hybrid Einstein-Debye model, this corresponds to an Einstein temperature 
of $\Theta_E\approx340$~K and a Debye temperature of $\Theta_D\approx270$~K 
taken from the density maximum at the upper cut-off of the optical modes.

\section{Results and Discussion}
\label{sec_rd}

\subsection{Crystal structure and sample quality}
\label{sec_xs}

\begin{figure}[h]
\centering
\includegraphics[width=\columnwidth]{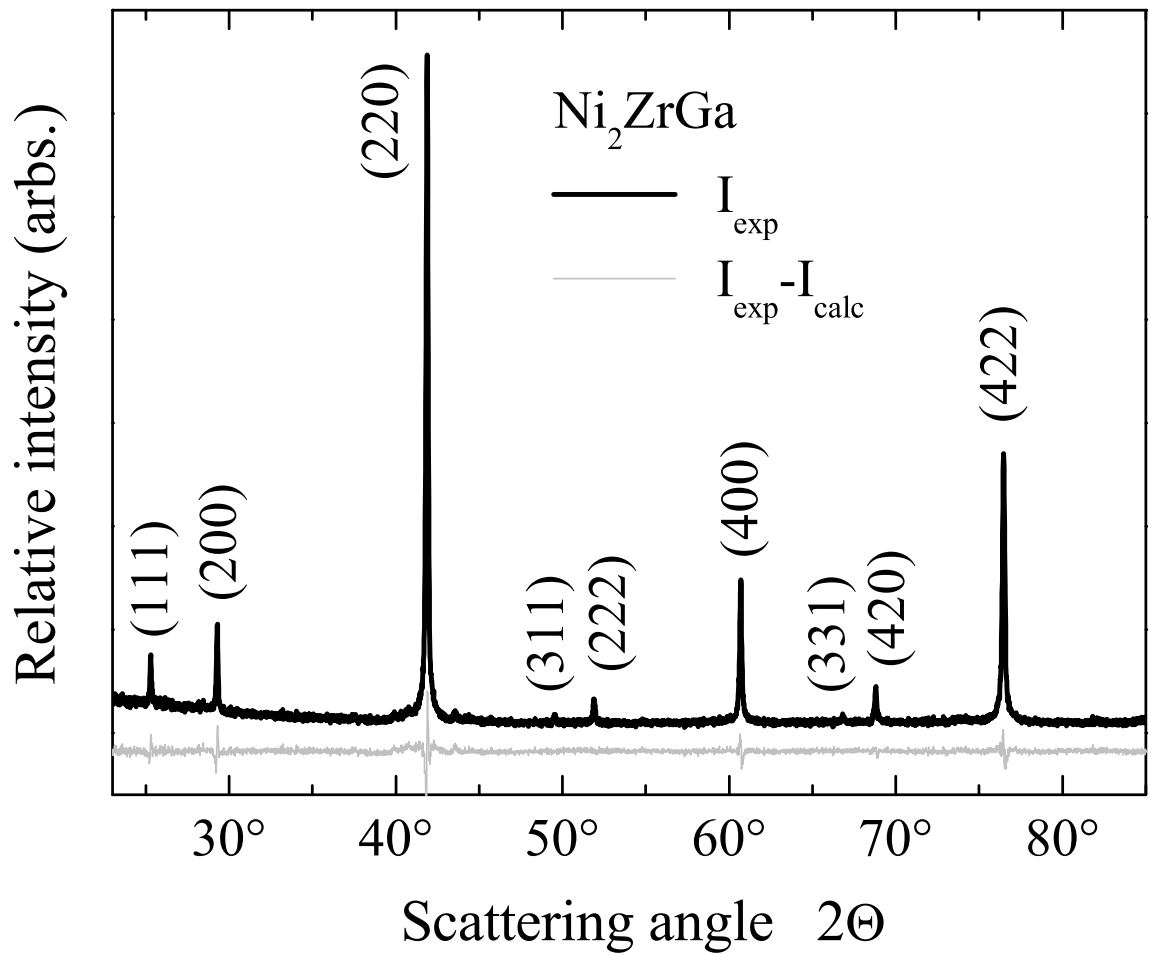}
\caption{Powder X-ray diffraction of ZrNi$_2$Ga at 300~K (black).
         The difference curve (grey) shows the difference between the
         observed data and the Rietveld refinement.}
\label{fig_xrd}
\end{figure}

ZrNi$_2$Ga crystallizes in the cubic L2$_1$ Heusler structure (space group:
$Fm\bar{3}m$), where the Wyckoff positions are 4$a$ (0,0,0) for Zr atoms, 4$b$ 
($\frac{1}{2}$,$\frac{1}{2}$,$\frac{1}{2}$) for Ga atoms, and 8$c$ 
($\frac{1}{4}$,$\frac{1}{4}$,$\frac{1}{4}$) for Ni atoms. Figure~\ref{fig_xrd} 
shows the diffraction pattern for ZrNi$_2$Ga with the raw data above (black) 
and the difference between a calculated Rietveld-refinement and the raw data 
below (grey). Within the experimental resolution of the diffractometer, 
no secondary phases were observed. The Rietveld refinement results
in a cubic lattice parameter of $a=6.098\pm0.003$~\AA. The as-cast 
samples of ZrNi$_2$Ga were indistinguishable from the annealed ones
in their XRD patterns, but magnetic, transport, and specific-heat measurements 
suggested an improved quality of the annealed samples. This improved quality
of the annealed crystals was confirmed by resistivity measurements yielding 
a residual resistivity ratio of two, which is typical for polycrystalline 
Heusler compounds. The specific-heat and magnetization measurements reveal sharp
superconducting transitions of $\Delta T_c/T_c \leq 0.03$. At low temperature, 
however, the measurements indicate small sample inhomogeneities or impurities.


\subsection{Properties of the superconducting state}
\label{sec_sc}

\begin{figure}[h]
\centering
\includegraphics[width=\columnwidth]{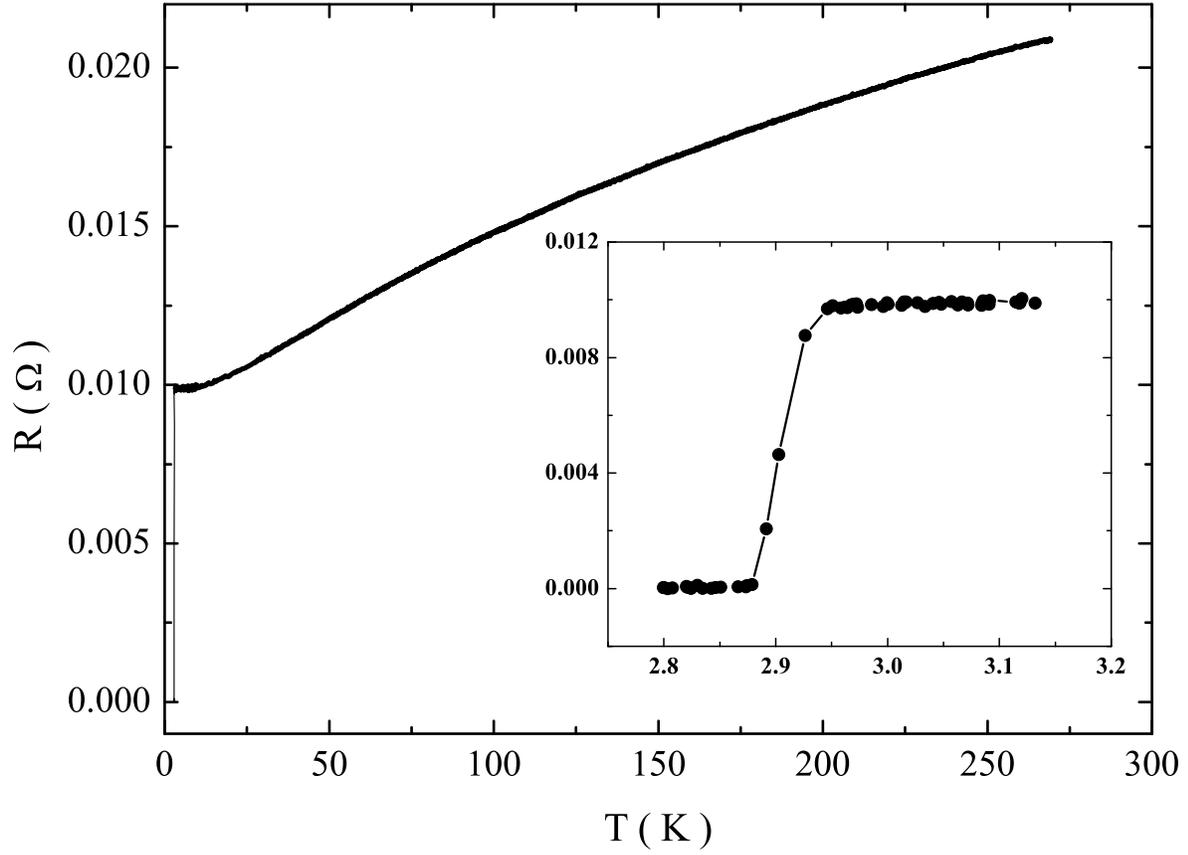}
\caption{The resistance of ZrNi$_2$Ga as a function of temperature. 
         The inset shows an enlargement of the superconduction transition
         at $T_c^{mid}=2.87$~K.}
\label{fig_RvT}
\end{figure}

The superconducting transition of ZrNi$_2$Ga was observed in measurements 
of the electrical resistance. Figure~\ref{fig_RvT} displays the temperature 
dependence of the resistance, which exhibits metallic behavior and a transition
to superconductivity at $T_c=2.87\pm0.03$~K. 

\begin{figure}[h]
\centering
\includegraphics[width=\columnwidth]{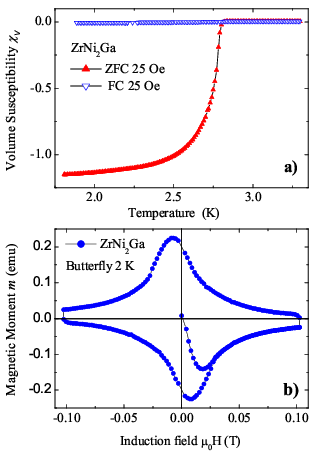}
\caption{ (Online in color) Magnetization measurements in the superconducting state of ZrNi$_2$Ga.
          Panel (a) shows the temperature dependent magnetization under ZFC
          and FC conditions. Panel (b) shows the field-dependent 
          magnetization at a temperature of 2~K.}
\label{fig_squid}
\end{figure}

Magnetization measurements using SQUID magnetometry were carried out 
to confirm bulk superconductivity in ZrNi$_2$Ga. The results of the magnetization 
measurements are given in Figure~\ref{fig_squid}. The upper panel (a) shows 
the temperature dependent magnetization $M(T)$ of a nearly spherical sample 
in an external field of $\mu_0H=2.5$~mT. A sharp onset of superconductivity
is observed in the ZFC curve at a temperature of $T_c=2.80$~K. The 
sharpness of the transition indicates good sample quality. The resisitive 
transition appears at a slightly higher temperature than tat determined from the
magnetization measurements. This is a well known phenomenon: the resistive
transition occurs when one percolation path through the sample becomes
superconducting whereas the magnetic transition requires a certain
superconducting volume. The ZFC curve demonstrates complete diamagnetic shielding.
For the calculation of the magnetic volume susceptibility, we used the
demagnetization factor $\frac{1}{3}$ of a sphere. The deviation from the 
expected value of -1 (100\% shielding) is ascribed to an imperfect spherical 
shape of the sample and therefore an underestimated demagnetization factor. 
The FC curve represents the Meissner effect for superconducting ZrNi$_2$Ga. 
The large difference between the ZFC and the FC curves shows clearly that 
ZrNi$_2$Ga is a type-II superconductor and points to a weak Meissner effect 
due to strong flux pinning. Figure~\ref{fig_squid}(b) shows a plot of the field 
dependent magnetization ($M$-$H$~curve). The magnetic field was varied from
-100~mT to 100~mT at a constant temperature of 2~K. The $M(H)$ measurements 
exhibit the typical butterfly loop of an irreversible type-II superconductor 
with large hysteresis due to strong flux pinning. An accurate determination 
of the lower critical magnetic field $H_{c1}$ at this temperature is nearly
not possible because of the broadening of the $M(H)$ curves. A very rough 
estimation of $H_{c1}$, defined as the magnetic field where the initial slope 
interacts with the extrapolation curve of $(M_{up}+M_{down})/2$, yields 
$\mu_0H_{c1}(T=2$~K) of approximately 16~mT compared to the upper critical 
field at $T=2$~K of 0.62~T.

\begin{figure}[h]
\centering
\includegraphics[width=\columnwidth]{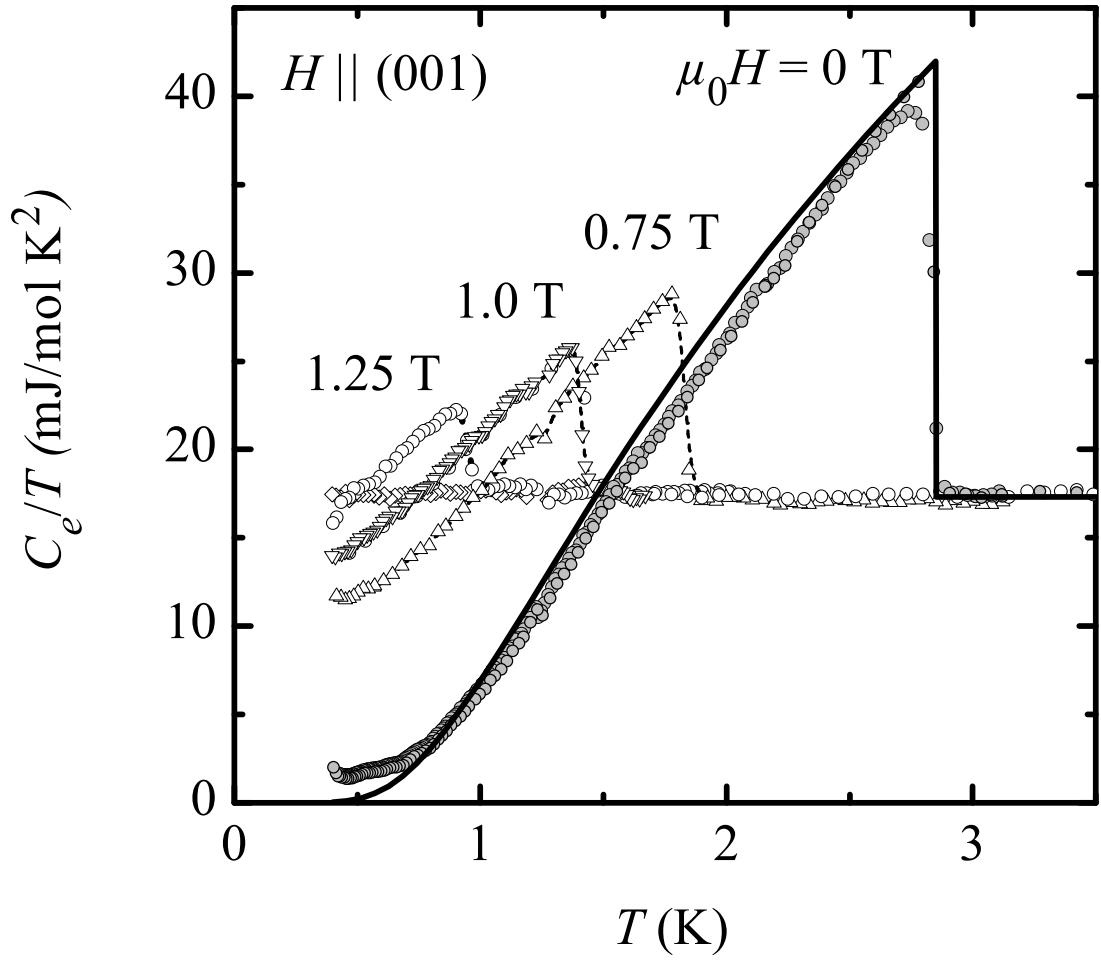}
\caption{Electronic contribution to the specific heat of
ZrNi$_2$Ga divided by temperature $T$ at various magnetic fields.
The continuous line represents the calculated behavior of a
weak-coupling BCS superconductor at zero magnetic field.}
\label{fig_ctb}
\end{figure}
\begin{figure}[h]
\centering
\includegraphics[width=\columnwidth]{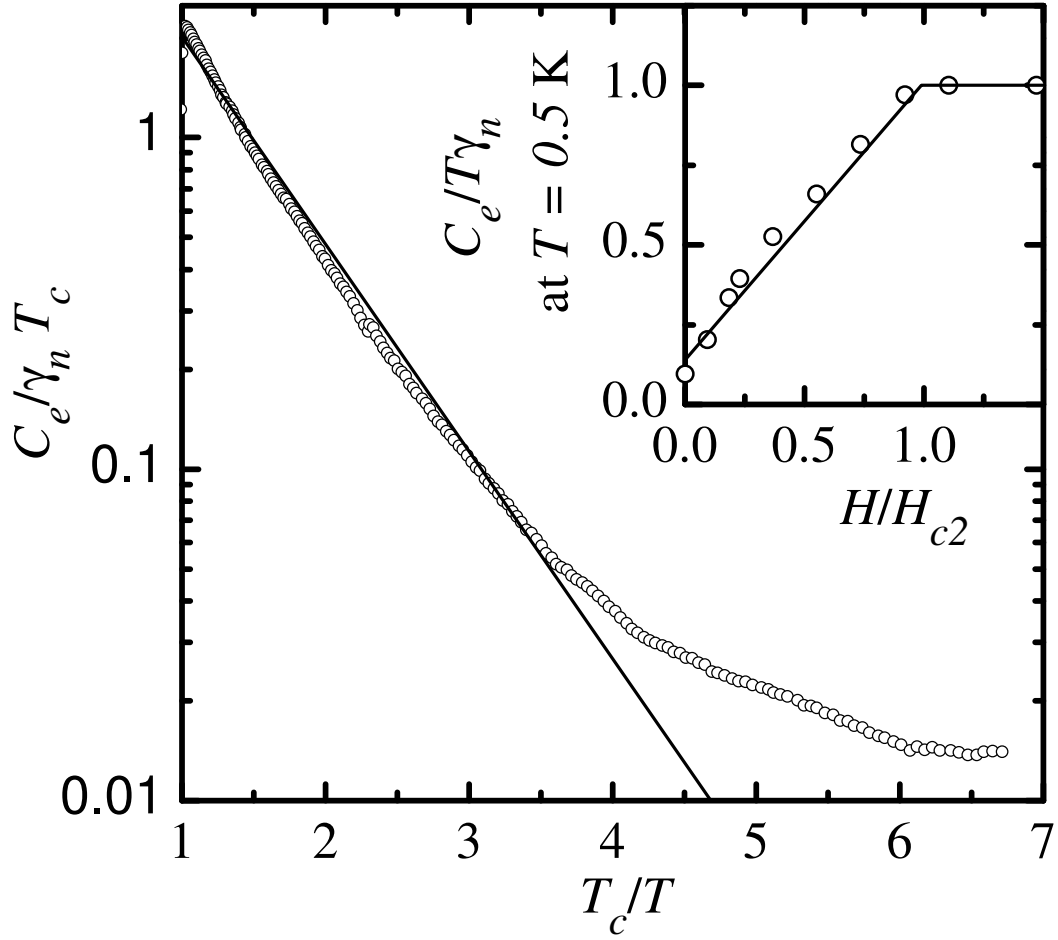}
\caption{Electronic contribution to the specific heat of
ZrNi$_2$Ga at zero field divided by $\gamma_n T_c/T$ vs. $T_c/T$. The inset
shows $C_e/T$ at $T=0.5$ versus the magnetic field.} 
\label{fig_ct}
\end{figure}

Figure~\ref{fig_ctb} shows the electronic contribution to the specific heat $C_e$ 
of ZrNi$_2$Ga plotted as $C_e/T$ vs. $T$ in various magnetic fields. The 
phonon contribution to the specific heat was subtracted as will be shown below. 
The mean feature of $C_e/T$ is the specific-heat jump $\Delta C_e$ at 
$T_c$ = 2.83~K with a width of 0.1~K. The nearly perfect agreement between 
the differently determined $T_c$ values together with the large $\Delta C_e$ 
confirm bulk superconductivity in ZrNi$_2$Ga. An analysis of the jump yields
$\Delta C_e/ \gamma_n T_c = 1.41$, which is in very good agreement with the
weak-coupling BCS value of 1.43. Here $\gamma_n$ denotes the normal-state 
Sommerfeld coefficient, which is discussed below. The energy gap is obtained 
from a plot of $C_e/\gamma T_c$ on a logarithmic scale versus $T_c/T$, as 
shown in Figure~\ref{fig_ct}. A comparison with the BCS formula for $C_e$ well 
below $T_c$

    \[C_e/\gamma T_c=8.5\exp[-(0.82\Delta(0)/k_BT]\]
    
yields an energy gap $\Delta(0)$ of 0.434~meV for $T \rightarrow0$ and 
$2\Delta(0)/k_B T_c=3.53$, again in very good agreement with the weak-coupling 
BCS value. At lowest temperatures one can observe deviations from the expected 
behavior. As these deviations are sample dependent and clearly reduced in the
annealed samples we attribute them to the aforementioned sample imperfections. 
In a more detailed analysis we compared $C_e$ at zero field with the calculated 
behavior of a BCS superconductor by using the approach of Padamsee $et~al.$~\cite{PNS73}
and the temperature dependence of the gap $\Delta(T)$ of M\"uhlschlegel~\cite{Mue59}.
In this model, $C_e$ is estimated for a system of independent fermion quasiparticles
with
\[
    \frac{S}{\gamma_n T_c} = -\frac{6}{\pi^2}\frac{\Delta(0)}{k_BT_c}\int^{\infty}_{0}[f\ln f+(1-f)\ln (1-f)]dy,
\]
\[
    \frac{C_e}{\gamma_n T_c} = t\frac{\partial (S/\gamma_n T_c)}{\partial t}
\]
where \[
f=[\exp(\sqrt{\epsilon^2+\Delta^2(t)})/k_BT + 1],~t=T/T_c,~y=\epsilon/\Delta_0.
\]
The only free parameter, the ratio $2\Delta(0)/k_BT_c$, was set to 3.53. 
Indeed, the specific heat can overall be rather well described by the weak-coupling 
BCS theory, as can be seen in Figure~\ref{fig_ctb}. To study the influence of the 
magnetic field we plot $C_e/T$ at a constant temperature of 0.5~K vs. the
$H/H_{c2}$ in the inset of Figure~\ref{fig_ct}. The linear increase of $C_e/T$ 
with $H$ corresponds to an isotropic gap, as expected for a cubic BCS superconductor.



\begin{figure}[h]
\centering
\includegraphics[width=\columnwidth]{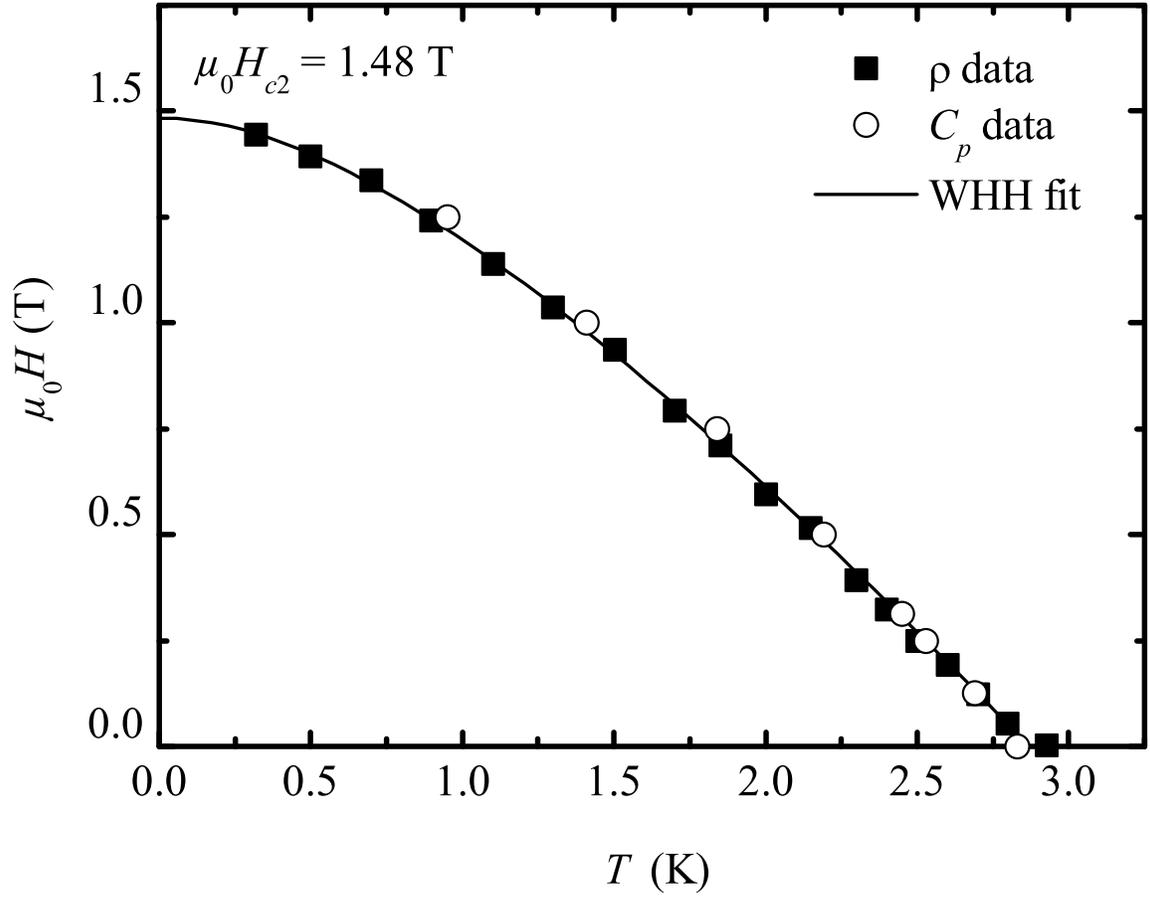}
\caption{Temperature dependence of the upper critical field
$H_{c2}$ of ZrNi$_2$Ga. Shown is a summary of the resistance and
specific-heat measurements. The continuous line represents a
calculation of the WHH model with $\alpha=0$ and $\lambda_{so}=0$,
which is identical to a finite $\alpha$ and
$\lambda_{so}\rightarrow \infty$.} 
\label{fig_Bc2}
\end{figure}

Further $R(T)$ measurements in various magnetic fields were performed to 
determine the upper critical field $H_{c2}$ of ZrNi$_2$Ga. In Figure~\ref{fig_Bc2} 
the data are summarized together with those of the specific-heat measurements. 
$H_{c2}(T)$ was theoretically derived by Wertheimer, Helfland, and Hohenberg
(WHH)~\cite{WHH66} in the limit of short electronic mean free path (dirty limit), 
including, apart from the usual orbital pair breaking, the effects of Pauli spin
paramagnetism and spin-orbit scattering. The model has two adjustable parameters: 
the Maki parameter $\alpha$, which represents the limitation of $H_{c2}$ by
the Pauli paramagnetism, and the spin-orbit scattering constant $\lambda_{so}$. 
$\alpha$ can be determined from the initial slope of the upper critical field
\[
 \alpha=-0.53\cdot\mu_0\left.dH_{c2}/dT\right|_{T=T_c} (\mu_0H \text{ in T}),
\]
or via the Sommerfeld coefficient $\gamma_n$ and the residual resistivity $\rho_0$ with:
\[
 \alpha=2e^2\hbar\gamma_n\rho_0/(2\pi^2 m k_B^2),
\]
where $m$ and $e$ are the free electron mass and charge, respectively. From the 
data we extract $\mu_0\left.dH_{c2}/dT\right|_{T=T_c}=-0.75$~T/K and $\alpha=0.4$. 
With $\lambda_{so} \rightarrow \infty$, the curve estimated by the WHH model 
follows the data points very closely, as is seen in Figure~\ref{fig_Bc2}. As 
the spin-orbit scattering counteracts the effect of the Pauli paramagnetism, this
is equal to $\alpha=0$ and $\lambda_{so}=0$, representing the upper bound of 
$H_{c2}$ where pair breaking is only induced by orbital fields. Consequently, 
the temperature dependence of $H_{c2}$ can either be explained by Pauli paramagnetism
with an extremely strong spin-orbit scattering or with a dominating orbital field 
effect. The critical field due to the Pauli term alone is 
$\mu_0 H_p(0)=\mu_0 \Delta(0)/\sqrt{2}\mu_B=1.84 T_c=5.24$~T, which is much higher
than $H_{c2}$ in the absence of Pauli paramagnetism
$\mu_0H_{c2}^*(0)=-0.69\cdot\mu_0\left.dH_{c2}/dT\right|_{T=T_c}=1.48$~T.
Hence pair breaking in ZrNi$_2$Ga is most probably only caused by
orbital fields~\cite{Clo62}. This is in contrast to other Ni-based Heusler 
superconductors like Ni$_2$NbGa and Ni$_2$NbSn where $H_{c2}^*(0)$ is clearly 
larger than the measured critical fields and therefore the Pauli paramagnetic
effect has to be considered (see Table~\ref{tab_heu}).

The thermodynamic critical field was calculated from the difference between the 
free energy of the superconducting and the normal states:
    \[
    \mu_0H_c=\left[2\mu_0\int^{T}_{T_c}\int^{T}_{T_c}(C_{e}/T''-\gamma_{n})dT''dT']\right]^{\frac{1}{2}}.
\]
A value of $\mu_0H_c=44.6$~mT is obtained. From the upper and thermodynamic critical
field one can estimate the Ginzburg-Landau parameter $\kappa_{GL}$, which is the 
ratio of the spatial variation length of the local magnetic field $\lambda_{GL}$ and the
coherence length $\xi_{GL}$: $\kappa_{GL}=H_{c2}(\sqrt{2}H_c)=\lambda_{GL}/\xi_{GL}=23.5$. 
The isotropic Ginsburg-Landau-Abrikosov-Gor'kov theory leads to the values of
$\xi_{GL}=\sqrt{\Phi_0/2\pi\mu_0H_{c2}}=15$~nm and $\lambda_{GL}=350$~nm 
($\Phi_0$ is the fluxoid quantum $h/2e$).

Obviously, ZrNi$_2$Ga is a conventional, weakly coupled, fully gapped type-II 
superconductor that is best described in terms of weak-coupling BCS superconductivity. 
If a phonon mediated pairing mechanism is assumed, we can determine the
dimensionless electron-phonon coupling constant $\lambda$ by using the McMillan 
relation~\cite{McM68}:
    \[
    T_c=\frac{\Theta_D}{1.45}\exp\left[\frac{-1.04(1+\lambda)}{\lambda-\mu_c^*(1+0.62\lambda} \right].
\]
If the Coulomb coupling constant $\mu_c^*$ is set to its usual value of 0.13 and 
$\Theta_D$ to our measured value of 300~K we get $\lambda= 0.551$, which is in good 
accordance with other superconducting Heusler compounds~\cite{WYM85}.



\subsection{Normal state properties}
\label{sec_ns}

Now we turn to a characterization of the normal state properties. When superconductivity 
is suppressed in a magnetic field of $H>H_{c2}$, the Sommerfeld coefficient $\gamma_n$ 
and the Debye temperature $\Theta_D$ can be extracted from the low-temperature behavior 
of the specific heat, $C=\gamma_nT + \frac{12}{5}\pi^4 Rn \theta_D^{-1}T^3$ where $R$
is the gas constant and $n$ the number of atoms per formula unit (= 4 in the case of
Heusler compounds). The extracted Debye temperature $\Theta_D=300$~K agrees very well 
with the calculated value of 270~K and is in the typical $\Theta_D$ range of other 
Heusler compounds (see Table~\ref{tab_heu}).

Likewise in accordance to our electronic structure calculations, the high density of 
states leads to a strongly enhanced Sommerfeld coefficient of 
$\gamma_n=\frac{\pi^2}{3}k_B^2N(\epsilon_F)=17.3$~mJ/mol~K$^2$. In fact, $\gamma_n$ 
is one of the highest values for paramagnetic Ni-based Heusler compounds 
(see Table~\ref{tab_heu}). As already stated by Boff $et~al.$~\cite{BFB96a}, the 
maximum of $\gamma_n$ in the isoelectronic sequence $A$ = Ti, Zr, Hf of $A$Ni$_2$C 
(C = Al, Sn) is found for Zr and in the sequence $A$ = V, Nb, Ta for V. As the 
electronic structure of all these compounds is quite similar, and consequently 
a rigid-band model may be applicable, the Fermi level can be shifted through the 
appropriate choice of $A$ to a maximum of $N(\epsilon_F)$~\cite{LiF91,BFB96a,BFB96b}.
This behavior and the comparatively large $\gamma_n$ of ZrNi$_2$Ga confirm
the van~Hove scenario.

\begin{figure}[h]
\centering
\includegraphics[width=\columnwidth]{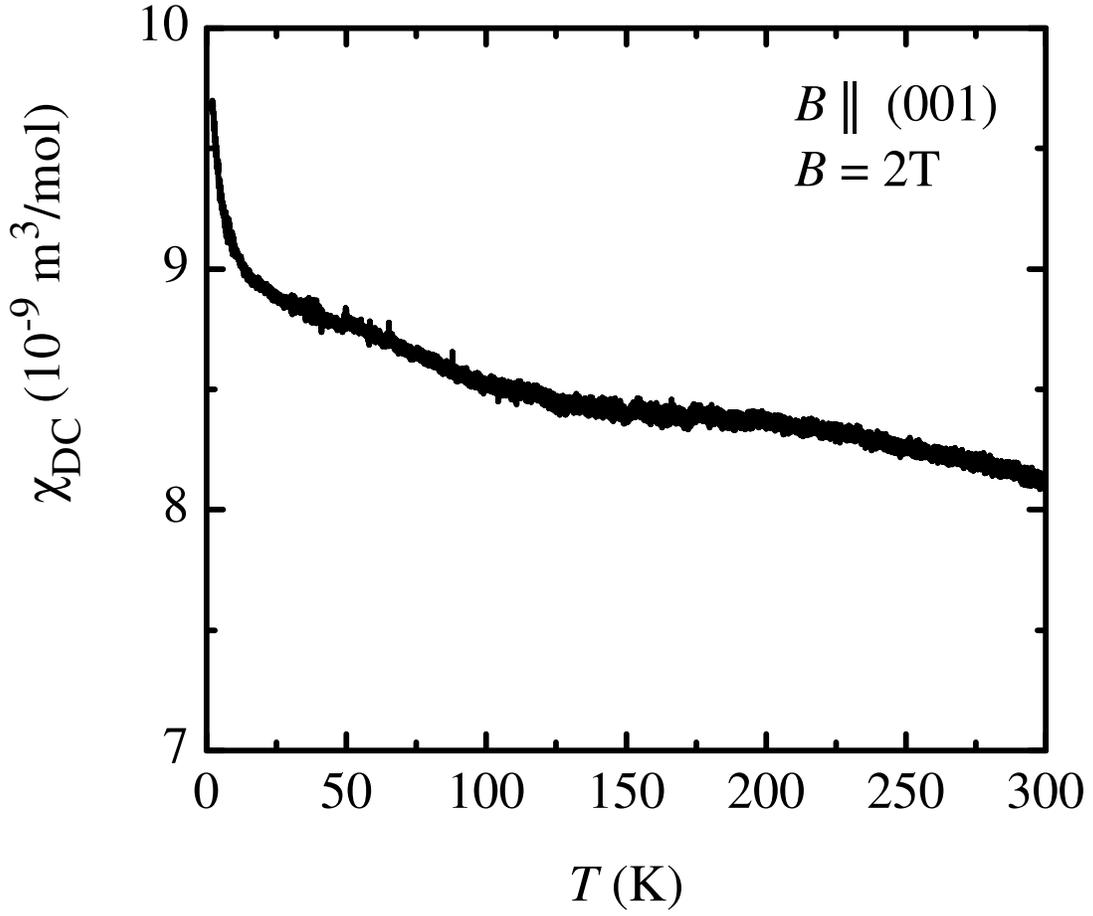}
\caption{Susceptibility $\chi_{DC}=M/H$ of ZrNi$_2$Ga in a
magnetic field of $\mu_0H=2$~T$>\mu_0H_{c2}$. The susceptibility
of the normal state shows Pauli-like behavior without any indications
of magnetic order. At low temperatures there is a small
Curie-Weiss-like upturn, which may be attributed to sample
inhomogeneities or impurities.} 
\label{fig_M}
\end{figure}

The measured magnetic susceptibility $\chi(T)$ as shown in Figure~\ref{fig_M} is 
nearly independent of $T$, indicative of a predominantly Pauli-like susceptibility. 
No sign of magnetic order can be found down to $T=1.8~$K. Even more, the low-temperature
specific-heat measurements demonstrate clearly that apart from the superconductivity 
no other phase transitions occur down to temperatures of 0.35~K. The enhanced 
susceptibility corresponds to the high density of states seen in $\gamma_n$ value 
as evidenced by the Wilson ratio $R = (\chi/\gamma_n) \cdot\pi^2 k_B^2/3 \mu_0 \mu_{eff}^2 = 0.97$ 
where we have set $\mu_{eff}^2=g^2\mu_B^2J(J+1)$ to its free electron values: i.e., 
the Land$\acute{e}$ factor $g=2$ and the total angular momentum $J=\frac{1}{2}$. 
The resulting Wilson ratio is close to that for independent electrons ($R = 1$).

Below about 10 K, a Curie-Weiss like increase of $\chi$ is observed for all samples. 
A fit of a Curie-Weiss law to the data yields a Weiss temperature of -3.3 K and an 
effective moment of 0.06\,$\mu_B$/f.u. (assuming $s = 1/2$). This Curie-Weiss like
behavior is sample dependent and can again be attributed to a small amount of magnetic
impurities. It is, however, supervising that no appreciable pair breaking is observed
as evidenced by the validity of the BCS law of corresponding states $2\Delta(0) = 3.53
k_BT_c$.

Finally, we want to discuss the influence of the increased DOS on the superconducting 
properties of ZrNi$_2$Ga. Although ZrNi$_2$Ga exhibits an enhanced $\gamma_n$ compared 
to the value $5.15$~mJ/mol~K$^2$ of NbNi$_2$Sn, both compounds have nearly the same 
transition temperature. Obviously, the simple relationship between $N(\epsilon_F)$ 
and $T_c$ does not hold. Table~\ref{tab_heu} demonstrates, likewise, that the upper
critical field $H_{c2}$ and the orbital limit $H_{c2}^*$ apparently do not depend on
the density of states in these materials. 

\section{Electron doping}
\label{sec_el}

The influence of the increased DOS on the superconducting properties is investigated from 
another point of view, which refers only to ZrNi$_2$Ga and the van~Hove singularity in
this compound: the Fermi level can be shifted with an appropriate choice of $A$, and the 
van~Hove scenario yields a maximum T$_c$ when the van~Hove singularity coincides with
$\epsilon_F$. According to the electronic structure calculations, electron-doping of 
ZrNi$_2$Ga should lead to this desired conicidence. Therefore, we doped ZrNi$_2$Ga with 
electrons in the $A$ position by substituting Zr with distinct amounts of Nb. The alloys 
Zr$_{1-x}$Nb$_x$Ni$_2$Ga with $x=0.15$, 0.3, 0.5, and 0.7 were prepared according to 
Section~\ref{sec_ed}.

\begin{figure}[h]
\centering
\includegraphics[width=\columnwidth]{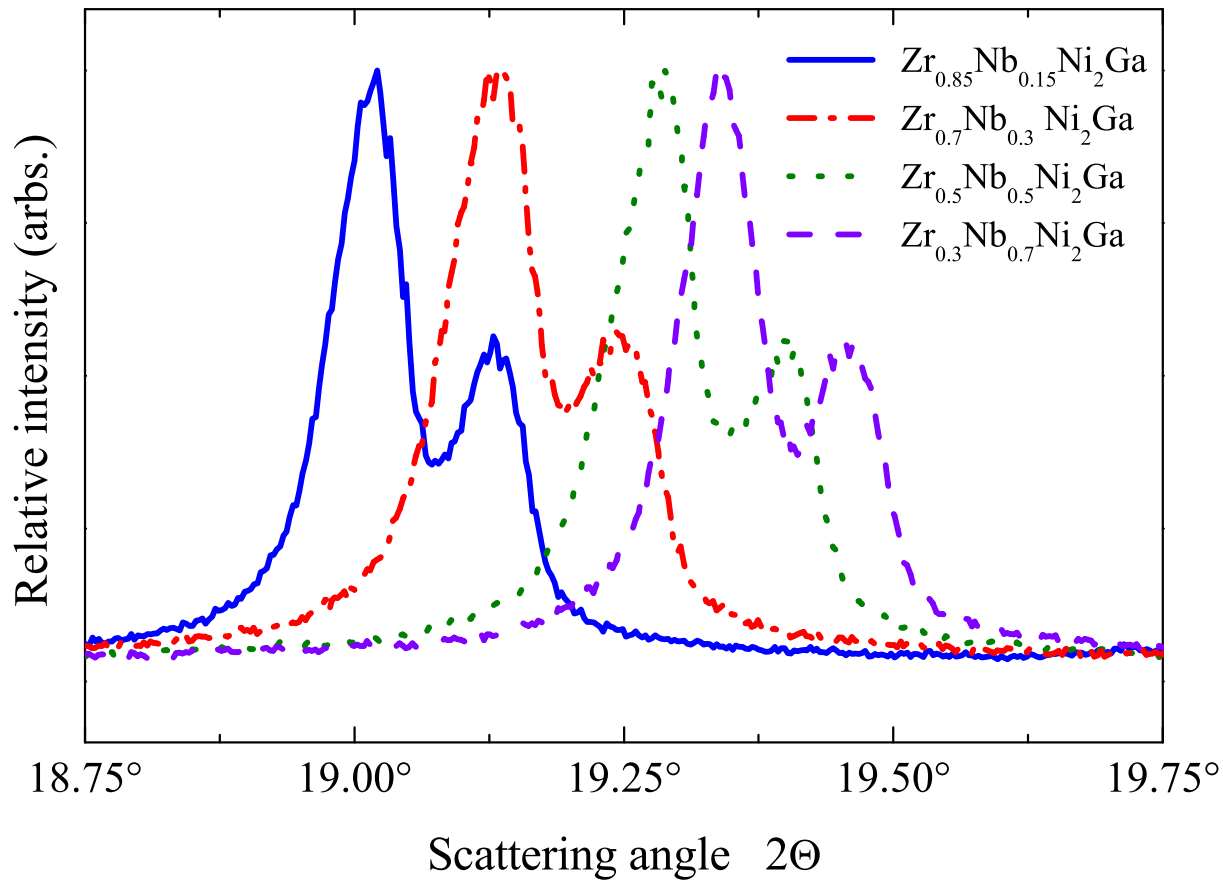}
\caption{Powder X-ray diffraction of the alloys Zr$_{1-x}$Nb$_x$Ni$_2$Ga at 300~K.
         Shown is the region around the (220) reflection, which determines the 
         cubic lattice parameter. The signals are splitted in Mo K$_{\alpha1}$
         and Mo K$_{\alpha2}$ peaks.}
\label{fig_dopxrd}
\end{figure}

The crystal structures of the alloys were determined using a Siemens D8 Advance diffractometer
with Mo K$_\alpha$ radiation. All alloys were found to crystallize in the Heusler structure 
(space group: $Fm\bar{3}m$). The atomic radius of Nb is smaller than the one of Zr, and 
thus a decrease of the lattice parameter is expected upon substituting Zr with Nb. In fact, 
this effect was observed (Figure~\ref{fig_dopxrd}). No impurity phases were detected in all 
alloys except of Zr$_{0.3}$Nb$_{0.7}$Ni$_2$Ga. The small difference between the lattice 
parameters of Zr$_{0.3}$Nb$_{0.7}$Ni$_2$Ga and Zr$_{0.5}$Nb$_{0.5}$Ni$_2$Ga supports that a 
saturation of Nb in the lattice of Zr$_{1-x}$Nb$_x$Ni$_2$Ga is reached for a value of 
0.5$\leq$x$\leq$0.7. Increasing the Nb concentration above the saturation limit leads to
a segregation of impurities. One of them was identified as elementary Zr.

\begin{figure}[h]
\centering
\includegraphics[width=\columnwidth]{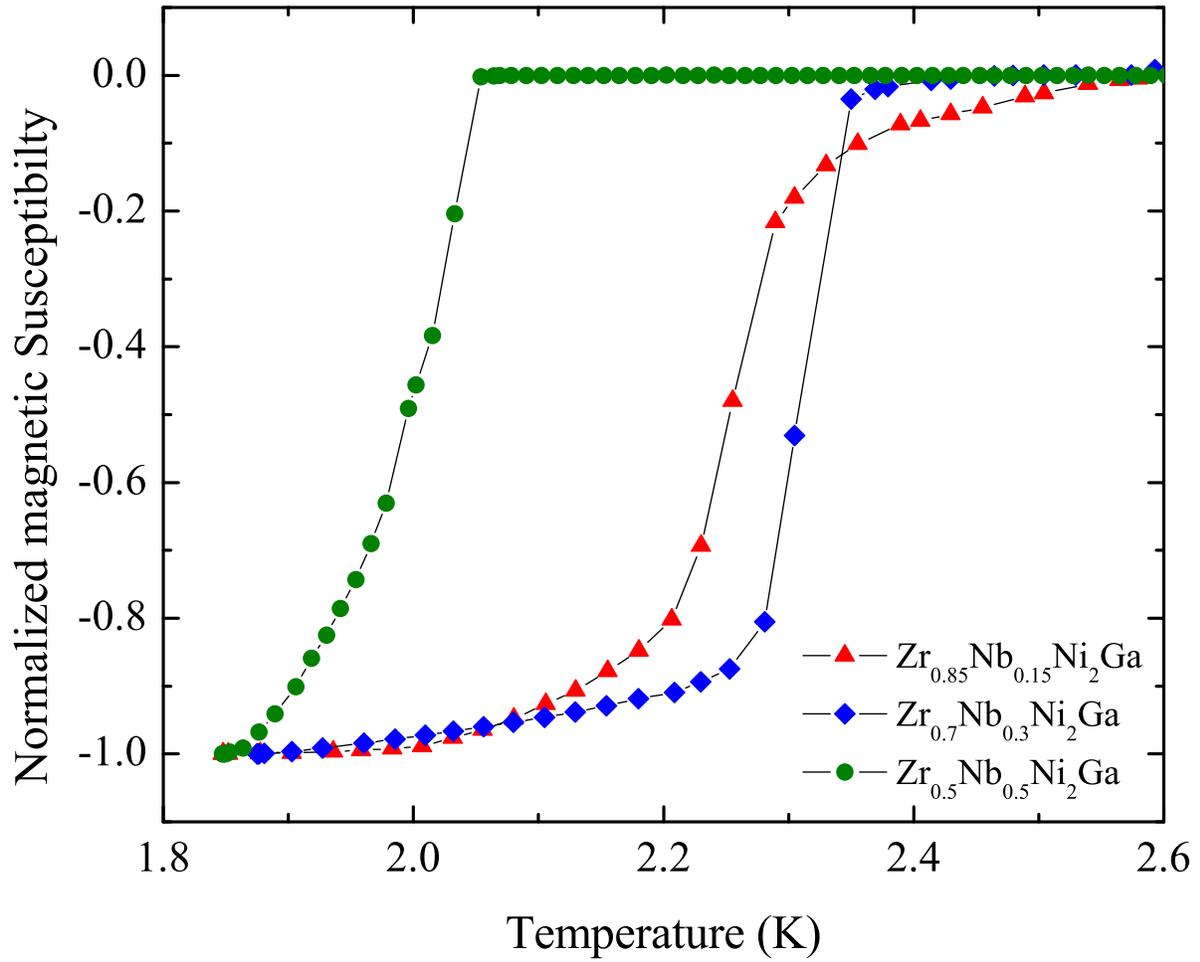}
\caption{ (Online in color) Superconducting transitions of the alloys Zr$_{1-x}$Nb$_x$Ni$_2$Ga under 
					ZFC conditions. The measurements were performed with magnetic fields of $\mu_0H=2.5$~mT,
					respectively.}
\label{fig_dopTc}
\end{figure}

The superconducting transitions of the alloys were analyzed in magnetization measurements using
a SQUID as described in Section~\ref{sec_ed}. Figure~\ref{fig_dopTc} shows the ZFC curves of the
alloys Zr$_{0.85}$Nb$_{0.15}$Ni$_2$Ga, Zr$_{0.7}$Nb$_{0.3}$Ni$_2$Ga, and Zr$_{0.5}$Nb$_{0.5}$Ni$_2$Ga.
Zr$_{0.3}$Nb$_{0.7}$Ni$_2$Ga did not show a superconducting transition down to 1.8~K. This is not
surprising because of the impurities, which were detected from XRD in this alloy. The other alloys show
a trend of decreasing T$_c$ with increasing Nb concentration as summarized in Table~\ref{Tab_sub}. 
It is therefore deduced that the Nb atoms act as additional scattering centres that suppress
the bulk superconductivity of ZrNi$_2$Ga.


\section{Conclusions}
\label{sec_co}

Starting with electronic structure calculations, the Heusler compound ZrNi$_2$Ga was predicted
to have an enhanced density of states at the Fermi energy $N(\epsilon_F)$ due to a van~Hove 
singularity close to $\epsilon_F$. According to the BCS model, ZrNi$_2$Ga was therefore 
expected to be an appropriate candidate for superconductivity with a comparatively high
superconducting transition temperature.

The predicted superconducting transition was found at $T_c$ = 2.87~K. Specific-heat
and magnetization measurements proved bulk superconductivity in this material and demonstrate 
that ZrNi$_2$Ga is a conventional, weakly coupled BCS type-II superconductor. The electronic
specific heat of the normal state shows a clearly enhanced Sommerfeld coefficient $\gamma_n$, 
which supports the van~Hove scenario. In the temperature range $0.35\text{~K} < T < 300~$K,
no sign of magnetic order is found. Apparently, the high $N(\epsilon_F)$ is not sufficient
to satisfy the Stoner criterion. The normal-state susceptibility is described best by an 
increased Pauli paramagnetism, corresponding to an enhanced $N(\epsilon_F)$. Despite the 
presence of magnetic impurities, which would suppress the energy gap by pair breaking, the BCS law
of corresponding states holds. This point deserves further investigations.

\bigskip
\begin{acknowledgments}
This work is funded by the DFG in Collaborative Research Center {\it "Condensed Matter
Systems with Variable Many-Body Interactions"} (Transregio SFB/TRR 49).
The work at Princeton was supported by the US Department of Energy division of
Basic Energy Sciences, grant DE-FG02-98ER45706. The authors would like to thank 
Gerhard Jakob for many suggestions and for fruitful discussions.
\end{acknowledgments}

\newpage


\newpage


\begin{table}[h]
\caption{\label{tab_heu} Comparison of nickel-based paramagnetic and superconducting 
Heusler compounds. Sommerfeld coefficient $\gamma_n$, Debye temperature $\Theta_D$, 
superconducting transition temperature $T_c$, orbital limit of the upper critical
field $\mu_0H_{c2}^*(0)=-0.69\cdot\mu_0\left.dH_{c2}/dT\right|_{T=T_c}$, and critical 
field $H_{c2}(0)$ extrapolated from low-temperature measurements.}
\begin{ruledtabular}
\begin{tabular}{lccccc}
 & $\gamma_n$ & $\Theta_D$ & $T_c$ & $\mu_0H_{c2}^*$ & $\mu_0H_{c2}$\\
 & (mJ/mol\,K$^2$) & (K) & (K) & (T) & (T)\\
\hline
ZrNi$_2$Ga & 17.3 & 300 & 2.85 & 1.48 & 1.48\\
TiNi$_2$Al & 13.37\footnotemark[3] & 411\footnotemark[3] & - & - & - \\
TiNi$_2$Sn & 6.86\footnotemark[4] & 290\footnotemark[4] & - & - & - \\

ZrNi$_2$Al & 13.67\footnotemark[3] & 276\footnotemark[3] & - & - & - \\

ZrNi$_2$Sn & 8.36\footnotemark[4] & 318\footnotemark[4] & - & - & - \\

HfNi$_2$Al & 10.85\footnotemark[3] & 287\footnotemark[3] & - & - & - \\
HfNi$_2$Sn & 6.37\footnotemark[4] & 280\footnotemark[4] & - & - & - \\

VNi$_2$Al & 14.17\footnotemark[3] & 358\footnotemark[3] & - & - & - \\

NbNi$_2$Al & 8.00\footnotemark[1],10.95\footnotemark[3] & 280\footnotemark[1],300\footnotemark[3] & 2.15\footnotemark[1] & 0.96\footnotemark[2] & $> 0.70$\footnotemark[1]\\
NbNi$_2$Ga & 6.50\footnotemark[1] & 240\footnotemark[1] & 1.54\footnotemark[1] & 0.67\footnotemark[2] & $\sim$ 0.60\footnotemark[1]\\
NbNi$_2$Sn & 4.0\footnotemark[1],5.15\footnotemark[3] & 206\footnotemark[1],208\footnotemark[3] & 2.90\footnotemark[1],3.40\footnotemark[3] & 0.78\footnotemark[2] & $\sim$ 0.63\footnotemark[1]\\

TaNi$_2$Al & 10.01\footnotemark[3] & 299\footnotemark[3] & - & - & -\\
\end{tabular}
\end{ruledtabular}
\footnotetext[1]{Ref.~\onlinecite{WYM85}}
\footnotetext[2]{calculated with the initial slope $dH_{c2}/dT$ from Ref.~\onlinecite{WYM85}}
\footnotetext[3]{Ref.~\onlinecite{RFB99}}
\footnotetext[4]{Ref.~\onlinecite{BFB96a}}
\end{table}

\begin{table}[h]
\caption{Properties of the alloys Zr$_{1-x}$Nb$_x$Ni$_2$Ga compared to ZrNi$_2$Ga. 
         $a$ are the measured lattice parameters, $T_c$ the critical temperatures from
         the ZFC curves in the magnetization measurements.}
\smallskip
\centering
\begin{tabular}{ l c c}
Compound/Alloy  									& $a$ (\AA) & $T_c$ (K) \\
\hline
ZrNi$_2$Ga 												& 6.098     &  2.8      \\
Zr$_{0.85}$Nb$_{0.15}$Ni$_2$Ga 		& 6.074     &  2.4      \\
Zr$_{0.7}$Nb$_{0.3}$Ni$_2$Ga			& 6.037   	&  2.3      \\          
Zr$_{0.5}$Nb$_{0.5}$Ni$_2$Ga			& 5.990     &  2.0      \\  
Zr$_{0.3}$Nb$_{0.7}$Ni$_2$Ga  		& 5.972     &  -        \\         
\hline
\end{tabular}
\label{Tab_sub}
\end{table}

\end{document}